\documentclass[article]{revtex4}
\usepackage{pdfpages}
\usepackage{epsf}
\usepackage{subfigure}
\usepackage{amsmath}
\usepackage{eqnarray,amsmath}
\usepackage{epstopdf}
\usepackage{mathrsfs}
\usepackage{natbib}
\usepackage{amssymb,latexsym}
\usepackage{dcolumn}
\usepackage{graphicx}
\usepackage{color}

\makeatletter
\newcommand*{\rom}[1]{\expandafter\@slowromancap\romannumeral #1@}
\makeatother
\def\be{\begin{equation}}
    \def\ee{\end{equation}}
\def\ba{\begin{eqnarray}}
    \def\ea{\end{eqnarray}}

\graphicspath{{images/}}
\begin{document}
    \title{\large \bf Nonlinear phenomena in general relativity }
    
\author{Alireza Allahyari}
\affiliation{Department of Physics, Sharif University of Technology,
    Tehran, Iran, and \\
    School of Astronomy, Institute for Research in Fundamental Sciences (IPM), P. O. Box 19395-5531, Tehran, Iran }
\email{allahyari@physics.sharif.edu}

\author{Javad T. Firouzjaee}
\affiliation{School of Astronomy, Institute for Research in Fundamental Sciences (IPM), P. O. Box 19395-5531, Tehran, Iran }
\email{j.taghizadeh.f@ipm.ir}
\author{Reza Mansouri}
\affiliation{Department of Physics, Sharif University of Technology,
    Tehran, Iran, and \\
    School of Astronomy, Institute for Research in Fundamental Sciences (IPM), P. O. Box 19395-5531, Tehran, Iran}
\email{mansouri@ipm.ir}
        
\begin{abstract}    
        
The perturbation theory plays an important role in studying structure formation in cosmology and post-Newtonian physics, but not all phenomena can be described by the linear perturbation theory. Thus, It is necessary to study exact solutions or higher order perturbations. Specifically, we study black hole (apparent) horizons and the cosmological event horizon formation in the perturbation theory. We emphasize that in the perturbative regime of the gravitational potential these horizons cannot form in the lower order. Studying the infinite plane metric, we show that to capture the cosmological constant effect we need at least a second order expansion.
        
\end{abstract}

    \maketitle
    \tableofcontents
\clearpage
\section{Introduction}

Recent advances in observational cosmology and astrophysics are providing a detailed testing of relativistic corrections to the large-scale structures and strong gravity regimes. In particular, the perturbative approach provides simple analyses to predict the cosmological and astrophysical data. Indeed, upcoming observational data (e.g. EHT \cite{EHT}, SKA \cite{SKA}), bring the opportunity to compare the exact relativistic models with the approximated perturbative ones. In the strong gravity side, especially in the black hole case, some features like black hole boundary \cite{cbh} and black hole mass \cite{cbh-mass} can not have well-defined meaning like the Newtonian ones. These features appear in the gravitational lensing \cite{cbh-lensing} or virialization \cite{cbh-virial} of the cosmological structures and gravitational waves physics \cite{gravitationalwave}. In the large-scale structure formation, there is some hints to go beyond the linear perturbation for the Friedmann-Lema\^itre-Robertson-Walker (FLRW) metric \cite{Creminelli:2004pv, Bertacca:2014dra,Baldauf:2011bh}. One important nonlinear effect is the coupling of short modes to long modes \cite{Allahyari:2017gmq}. The other important nonlinear effect is the non-Gaussianty which can be used to discriminate the inflationary models. However, this needs a second-order expansion of the inflationary potentials \cite{Acquaviva:2002ud}.\\

One hint to go beyond the linear order is that if we add a tensorial spin-two fields on the Minkowski background and couple this field to its energy-momentum iteratively infinite times, it is shown that one recovers the Einstein-Hilbert action \cite{Deser:1969wk}. This implies that there are phenomena that a perturbative study may not capture at finite order.\\

One method to study Einstein equations when we do not have the exact solution is the perturbation theory. In our universe, because the initial deviations from the FRW metric are assumed to be small, one is justified to consider these deviations as perturbations \cite{bardeen}. To increase the precision, one can go to higher orders in a perturbative expansion. Higher-order perturbation theory is discussed in the literature \cite{Bruni:1996im,Nakamura:2004wr}. There are phenomena where the perturbative solutions do not approximate the exact results. One example is the exact lensing of dynamical structures  studied in  \cite{cbh-lensing}.\\

Our aim in this paper is to study the phenomena that need a nonlinear treatment like black hole horizons and the cosmological horizon. In the case of the primordial black hole, It was shown the black hole horizon cannot form in the linear phase of the cosmological perturbations \cite{pbh-per}. Specifically, we study the horizons in the Schwarzschild and LTB metrics. To study the horizons, we use the concept of the apparent horizon. Treating the Schwarzschild metric as a perturbation on the Minkowski metric, we show that its apparent horizon appears at the fourth order in the expansion of the gravitational potentials. We can extend this study to the dynamical spherically symmetric models. In this case, we investigate whether black hole apparent horizon forms in terms of Minser-Sharp gravitational potential. \\

The de Sitter metric is the basic model for the accelerated expansion of the early and late universe. This metric has a horizon defined by the value of the cosmological constant. An observer can not receive any signal beyond this horizon. Note that even the presence of a small  cosmological constant leads to crucial differences \cite{Ashtekar:2017dlf}. We also write this metric as perturbations on the Minkowski metric. One finds that the cosmological horizon appears in the fourth order. There is a way to treat the de Sitter metric as a perturbation on the Minkowski metric as a first-order approximation and extend this to all orders. This yields the de Sitter metric in the Kerr-Schild form. The perturbative regime of the de Sitter metric is characterized by the cosmological constant potential term.\\

To show the especial case where the effect of the cosmological constant appears at nonlinear order, we study the metric of an infinite plane. We choose this metric because it describes a quasi-local structure. Note that the exact solution for this metric requires a cosmological constant. We try to approximate this metric by expanding the gravitational potentials. Because the Einstein tensor vanishes to first order, this implies that the cosmological constant effect appears at higher orders. \\

This paper is organized as follows. In section II, we study the Schwarzschild metric and LTB metric. Using the apparent horizon, we show that the horizon appears in fourth order.  Following the discussion about the black hole, we will show that the apparent horizon for a dynamical spherically symmetric black hole can not form in the perturbation of the Misner-Sharp mass potential. In section  III, we extend the study of nonlinear phenomena to the cosmological models where we show that the de Sitter horizon would require a fourth order expansion. In section IV, we study the infinite plane metric that describes a quasi-local structure. We show that the effect of the cosmological constant appears in second order. We conclude in section V.\\

\section{Black hole formation}
In general, relativity, if the energy conditions are satisfied, and a collapsed object compacts into a high-density matter which the light trapped in a region around it, the black holes form. Black holes are defined by their event horizon. The event horizon is a global definition. Alternatively, it is more convenient to define a black hole in terms of local quantities such as an apparent horizon \cite{Nakamura}. The apparent horizon is defined as the surface on which the convergence of the outgoing null geodesics $\theta$ vanishes \cite{Ashtekar:2004cn}. In this section, we study the black hole apparent horizon for a perturbation of the Minkowski spacetime. Specifically, we seek to find the order in which the apparent horizon appears. 

\subsection{Stationary Black hole formation}

Let us consider a spherically symmetric metric written as linear perturbations on the Minkowski metric. The general form of this metric is given by \cite{Harada:2015yda}
\begin{align}
    ds^2=(\eta_{\mu\nu}+h_{\mu\nu})dx^{\mu}dx^{\nu}=(-1+A(r,t))dt^2+(1+B(r,t))dr^2+C(r,t)dtdr+r^2(1+D(r,t))d\Omega^2.
    \label{eq:1}
\end{align}
The quantity which characterizes a black hole is the expansion of null geodesics $\theta=\nabla_{\mu}k^{\mu}$, where $k^{\mu}$ is the photon null vector. The surface on which the outgoing expansion is zero is called the apparent horizon. 

Using the metric in Eq. (\ref{eq:1}) yields
\begin{equation}
    \theta=\sqrt{2}(1-D/2)r\left(r\partial_{t}D/2+(1-B/2-C)((1+D/2)+r\partial_{r}D/2) \right) .
\end{equation}
Since the $ (D,B,C) $ are the perturbative quantities $ (D,B,C)  \ll 1 $, it can be seen that the expansion cannot be zero in the perturbative regime. As a result, the apparent horizons can not form in the models which are approximated with this metric.\\

To know whether the horizon appears by going to nonlinear orders, we study a perturbative solution which is static and has s spherical symmetry. We also suppose that metric is in isotropic coordinates. Thus, this metric can be written as
\begin{align}
    ds^2=(-1+f(r))dt^2+(1+g(r))(dr^2+r^2d\Omega^2),
\end{align}
where  $r$ coordinate is related with area coordinate with $\sqrt{(1+g(r))}r=R$. We expand the metric to fourth order  $f(r)=\sum_{n=1}^{4} f^{(n)}$ and we define $g(r)=\sum_{n=1}^{4} g^{(n)}$ \cite{Ngubelanga:2015dih}. 
We solve Einstein equations order by order. This yields
\begin{align}
    f(r)=2 m /r - 2 m^2 /r^2 + 3 m^3/2 r^3 - m^4/r^4\\
    g(r)=2 m /r + 3 m^2 /2 r^2 + m^3 /2 r^3 + m^4 /16 r^4.
\end{align}
Now we study the apparent horizon for this metric. In the exact model the apparent horizon is located in $R=2m=4r$.
In Isotropic coordinates to fourth order, we obtain  the expression for the apparent horizon as
\begin{equation}
    \theta=2(\frac{1}{r}+\frac{4 M  }{r^2}+\frac{11 M^2 }{2 r^3}-\frac{165 M^4 }{16 r^5}+O\left(5\right)).
\end{equation}
We see that the expansion vanishes at the fourth order leading to the apparent horizon. This perturbative expansion is equivalent to expansion in terms of the gravitational potential $x=\frac{2m}{R} <1$. It can be seen that the apparent horizon can form in the regime that the perturbation holds i.e $ x\sim 0.5 $.\\

\subsection{Dynamical Black hole formation}

The spherically symmetric exact solution of Einstein equations with dust is given by the LTB metric. This metric is given by
\begin{equation}
ds^2=-dt^2+\frac{(\partial_{r} R(t,r))^2}{1+2f(r)}dr^2+R(t,r)^2 d\Omega^2
\end{equation}
where $f(r)$ is the energy for each spherical shell with radius $r$.
The Einstein equations for this metric lead to
\begin{align} \label{ein-eq}
\dot{R}(t,r)=2M(r)/R+2f(r)\\
\rho=\frac{\partial_{r}M(r)}{R(t,r)^2\partial_{r}R(t,r)},
\end{align}
where dot means derivative with respect to time.
We can have three classes of solutions depending on the curvature function, $f(r)$. $M(r)$ is the Misner-Sharp mass that can be applied to the cosmological structure mass profiles \cite{cbh-mass}. 
In this metric the surface where $R(t,r)=2M(r)$ is the apparent horizon.

\begin{figure}[htbp!]
    \centering
    \includegraphics[width=0.6\columnwidth]{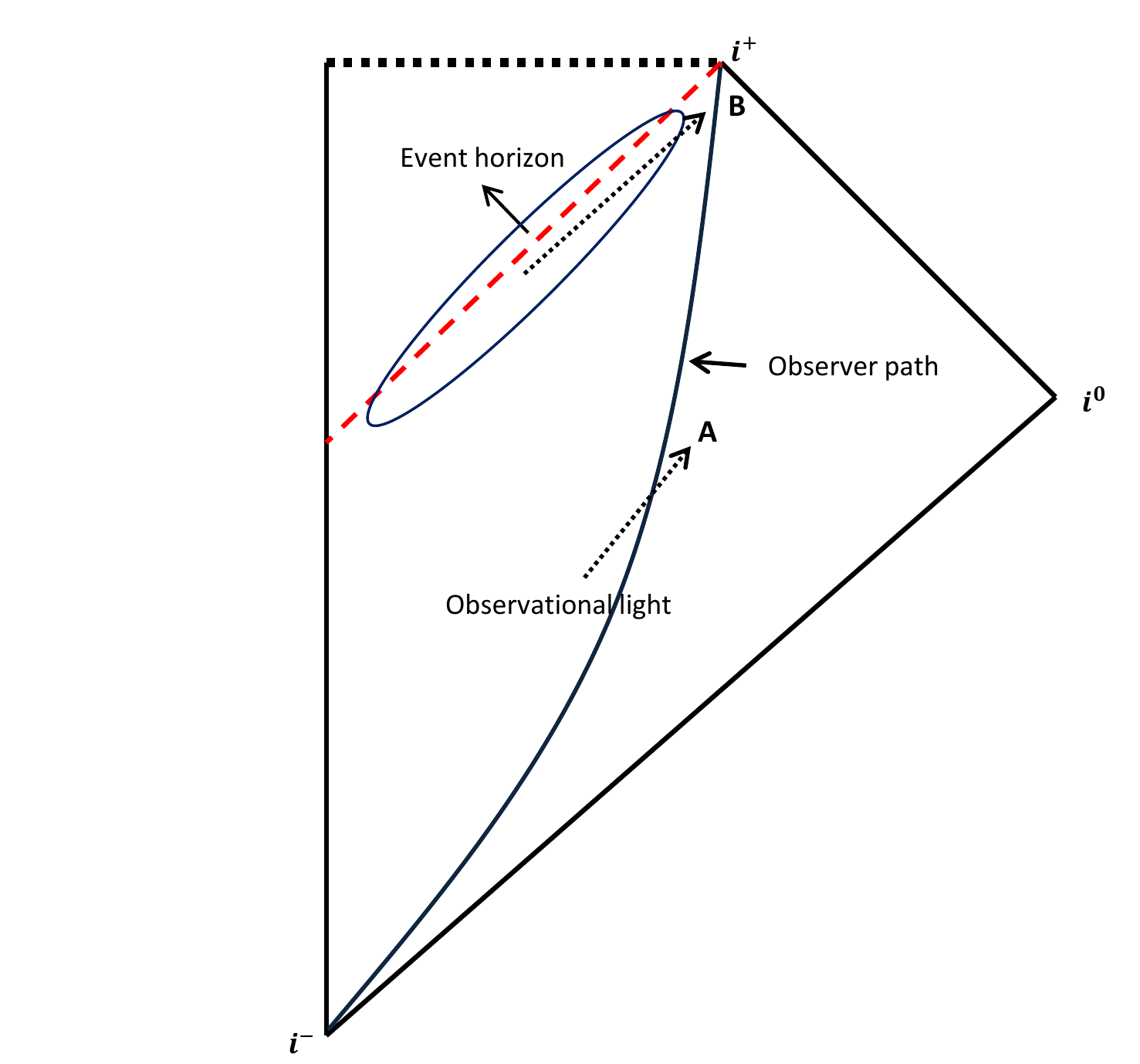}
    \caption{This Penrose diagram schematically describes the region (in the oval) where perturbation can not have a correct description of the physical phenomenon.  B observer which receives a light signal from the physical events in this region (near the black hole horizon) can not apply the perturbative approach in contrast to  A observer.}
    \label{bhpenrose}    
\end{figure}

One can show that the metric is also nonlinear at the time of the apparent horizon formation. First we transform the metric to a new coordinate $(t,R)$ given by
\begin{equation}
ds^2=-(\frac{1-\frac{2M}{R}}{1+2f})dt^2+\frac{dR^2}{1+2f}-2\frac{\sqrt{\frac{2M}{R}+2f}}{1+2f}dRdt+R^2 d\Omega^2.
\end{equation}
Here, we used the Einstein
equation (\ref{ein-eq}) to simplify the metric. If we interpret $ \Phi= \frac{M}{R} $ as the gravitational potential, in the regime that the gravitational potential tends to zero and the spacetime becomes flat, $ f(r)=0 $, this metric asymptotes to the Minkowski metric at infinity in these coordinates.
We  see that  the apparent horizon forms at $g_{tt}=0 $ which gives $R(t,r)=2M(r)$. It is clear from the metric that as long as the gravitational potential of dust is in the linear regime $ 2\Phi < 1 $,   the apparent horizon can not form. Thus, the gravitational potential which is defined by the Misner-Sharp mass is a good criterion to present a black hole existence which is a nonlinear feature of the collapse. This analysis can be extended for the black hole formation of a perfect fluid collapse \cite{Moradi:2013gf} and we can see that in the linear regime $ 2\Phi < 1 $,  the apparent horizon can not form.
The non-perturbative region is schematically shown in Fig. (\ref{bhpenrose}).\\

\section{Cosmological Horizon}

Similar to the black holes, there is a boundary beyond which an observer (in a cosmological spacetime that is expanding)  cannot receive any signal. Similar to the previous section, we study the cosmological event horizon. We write the metric as a perturbation on the Minkowski spacetime. Specifically, we seek to find the order in which the cosmological event horizon appears.

\subsection{de Sitter metric}
In this part, we study the de Sitter metric that is a cosmological metric. This metric is the basic model of the early inflationary phase and late time acceleration in cosmology \cite{Faraoni:2000nt}. The important feature of this metric is that it has a cosmological horizon.
Similar to the Schwarzschild metric, we write the de Sitter metric in isotropic coordinates as
\begin{equation}
ds^2=-(1+H^2 r^2/4)^{-2}(1-H^2 r^2/4)^2 dT^2+\frac{1}{(1+H^2 r^2/4)^2}(dr^2+r^2d\Omega^2)
\label{met1}.
\end{equation}

 If we solve the Einstein equations with a cosmological constant in the vacuum up to the fourth order of cosmological constant potential, $ (1 > H^2 r^2/4) $, in the isotropic coordinates we get
\begin{eqnarray}  
ds^2 &=& (-1 + 3/4 H^2 r^2 - 5/16 H^4 r^4  +
7/64 H^6 r^6  - 9/256 H^8 r^8 )dT^2 \nonumber \\ &+&
(1 - 1/2 H^2 r^2  + 3/16 H^4 r^4  - 
1/16 H^6 r^6  + 5/256 H^8 r^8 )(dr^2+r^2d\Omega^2).
\end{eqnarray}
One can show that the cosmological event horizon will form in the region that the expansion of the ingoing null geodesic becomes zero. This is equivalent to $g^{rr}=0$ \cite{Hayward:1993wb}. We find that similar to the Schwarzschild metric, the horizon appears at fourth order.\\

Let us study an observable quantity. For cosmological observations, the metric is written  in the comoving coordinates as

$$ds^2=-dt^2+e^{2H t}(dR^2+R^2 d\Omega^2).$$

One observable in  a cosmological setting is the luminosity distance. The luminosity distance in a de Sitter universe can be given by
\begin{equation}
d_{L}=(1+z) d_{A}=e^{-H t_{s}}(\frac{1}{H})(e^{-H t_{s}}-1),
\end{equation}
where $t_{s}$ is the source physical time in the comoving coordinates. In the limit that  $t_{s}$ is small we find the luminosity distance in the Minkowski metric, $ d_{L}= d_{A}=t_s $. 


It can be inferred that near the horizon $t_{s}\sim 1/H$, the luminosity distance could not be written as a perturbation over a Minkowski metric.

\begin{figure}[htbp!]
    \centering
    \includegraphics[width=0.5 \columnwidth]{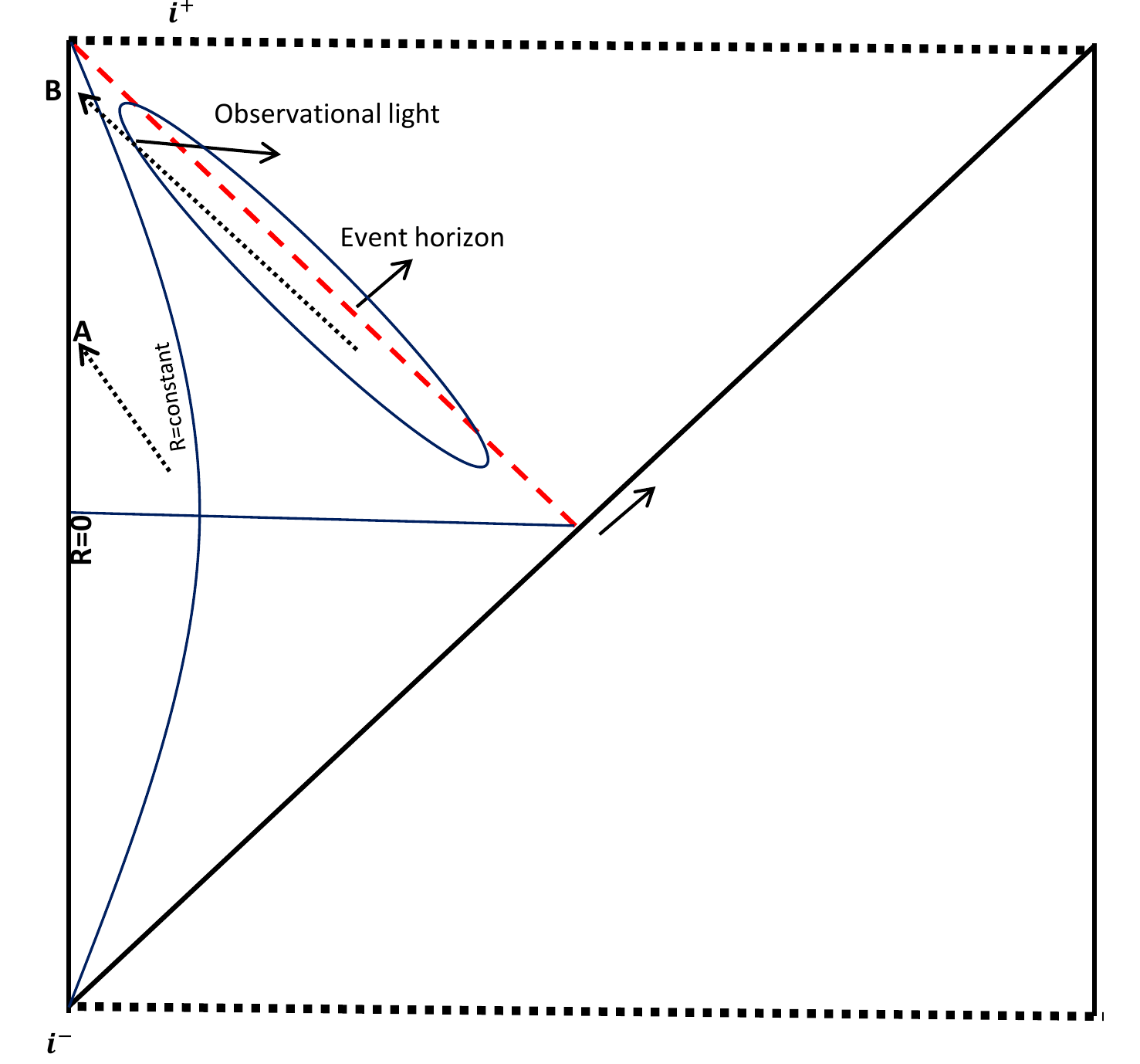}
    \caption{This Penrose diagram of the de Sitter spacetime schematically describes the region (in the oval) this the background perturbation can not have a correct description of the physical phenomenon in that region. The B observer which receives a light from the physical events in this region (near the de Sitter horizon) can not apply the perturbative approach in contrast to the A observer. This shows that the more large-scale observational data give the less validity of the perturbative approach.}
    \label{co-penrose}    
\end{figure}

The non-perturbative region is schematically shown in Fig. (\ref{co-penrose}).\\

\subsection{de Sitter in Kerr-Schild form} 
There is a way to write the de Sitter metric as a perturbation on the  Minkowski spacetime for scales smaller than the Hubble scale and extend this to nonlinear orders.  We start with the Minkowski metric with one null coordinate $u$ and add a first-order perturbation $h$ in $u$ direction. The metric is given by
\begin{align}
    ds^2=-2drdu-du^2+r^2d\Omega^2+h du^2.
\end{align}
To first order in the cosmological constant potential,  Einstein equations yield $h=1/3  H^2r^2$.
A closer inspection of the Einstein equations shows that this metric is the nonlinear solution  \cite{Harte:2016vwo}. This metric is actually the exact form of de Sitter metric.

Though the exact de Sitter solution (nonlinear solution) can be written in the linear perturbative form as the metric above, the real perturbation parameter appears in $h$ which presents the cosmological constant potential.\\


\section{Infinite plane metric} 
In this section, we study a vacuum solution of the Einstein equations with planar symmetry with the cosmological constant. The metric with planar symmetry  we consider is given by
\begin{equation}
    ds^2=-e^{2g(z)}dt^2+e^{2f(z)}(dx^2+dy^2)+dz^2.
    \label{plane}
\end{equation}
It is important to note that the vacuum solution of the Einstein's equations  does not exist for this metric, unless we have a cosmological constant of the form $\lambda=3 a^2/4$ where $a$ is related to the surface mass density of the plane. The solution yields  $f(z)=g(z)=a\lvert z \lvert$.\\

We write this metric as an expansion around Minkowski metric as

\begin{equation}
    ds^2=(-\sum a_{i}z^{i})dt^2+(\sum b_{i} z^{i})(dx^2+dy^2)+dz^2,
\end{equation} 
where we set $a_{0}=b_{0}=1$.
We solve the Einstein equations without a cosmological order by order.  We find that the first order perturbation is always a solution. Consequently, the first order perturbation does not allow for a cosmological constant. In other words, the effect of the cosmological constant for this metric  should appear at higher orders.

If we solve the Einstein equations at second order with a cosmological constant term, we need to have $a_{1}=b_{1}$. This example shows that the effect of the cosmological constant appears at second order.\\

\section{Conclusion}
Although, it is common in cosmology and gravity to assume that  linear perturbation theory is needed for a sufficient description of
clustering of Large-Scale Structures and gravitational waves physics, recently there are some attempts that show the nonlinear physics appears in our observations \cite{Baldauf:2011bh,Bertacca:2014dra,Bruni:1996im,Creminelli:2004pv,cbh-mass,cbh,gravitationalwave}. We study phenomena that need a nonlinear treatment in gravity. We show that the apparent horizon in a spherically symmetric stationary black hole solution appears at the fourth order in perturbations of the gravitational  potential.  Studying the spherically symmetric dynamical black holes have shown that the gravitational potential which is defined by the Misner-Sharp mass is a good criterion to present a black hole existence which is a nonlinear feature of the collapse.

The other case we study is the cosmological event horizon in de Sitter metric where we show that the horizon appears in the fourth order in the perturbations of the gravitational potential. Writing this metric in the Kerr-Schild form, we find a linear solution that is also the exact solution. The perturbative regime of the de Sitter metric is characterized by the $h$ term which represents the cosmological constant potential. 

The other phenomena which we consider are the effect of the cosmological constant in perturbation theory. Thus, we adapt the infinite plane solution which is a vacuum solution with a cosmological constant. We show that the effect of the cosmological constant, in this case, can be captured at second order in terms of the gravitational potential. Our study shows that black hole apparent horizons, the cosmological event horizon and the cosmological constant in the infinite plane solution are features that appear in nonlinear physics.\\



\clearpage

\end{document}